\documentclass[onecol]{epl2}
\title{Slow optical solitons via intersubband transitions in a semiconductor quantum well}

\author{Wen-Xing Yang\inst{1,2} \and Ray-Kuang Lee\inst{1}}

\institute{\inst{1} Institute of Photonics Technologies, National
Tsing-Hua University, Hsinchu 300,
Taiwan\\
\inst{2} Department of Physics, Southeast University, Nanjing
210096, China}




\abstract{We show the formation of bright and dark slow optical
solitons based on intersubband transitions in a semiconductor
quantum well (SQW). Using the coupled Schr\"odinger-Maxwell
approach, we provide both analytical and numerical results. Such a
nonlinear optical process may be used for the control technology of
optical delay lines and optical buffers in the SQW solid-state
system. With appropriate parameters, we also show the generation of
a large cross-phase modulation (XPM). Since the the intersubband
energy level can be easily tuned by an external bias voltage, the
present investigation may open the possibility for electrically
controlled phase modulator in the solid-state system.}

\begin{document}

\maketitle

Solitons describe a class of fascinating shaping-preserving wave
propagation phenomena in nonlinear media. Over the past few years,
the subject of extensive theoretical and experimental investigations
on solitons in optical fibers\cite{1, 2}, cold-atom media\cite{3, 4,
5, 6, 7}, Bose-Einstein condensates (BEC)\cite{8, 9}, and other
nonlinear media\cite{10}, has received a great deal of attentions
mainly due to that these special types of wave packets are formed as
the result of interplay between nonlinearity and dispersion
properties of medium under excitations, and can lead to undistorted
propagation over extended distance. In the optical domain, most
optical solitons are produced with intense electromagnetic fields,
and far-off resonance excitation schemes are generally employed in
order to avoid unmanageable optical field attenuation and distortion
\cite{1}. As a result, optical solitons produced in this way
generally travel with a propagation speed very close to the speed of
light in vacuum. As well known, the wave propagation velocity in
highly resonant medium can be significant reduced via
electromagnetically induced transparency (EIT) technique \cite{11}
or Raman-assisted interference effects. Recently, ultraslow optical
solitons including two-color solitons with very low group velocities
based on EIT technique or Raman-assisted interference effects, have
been studied in atomic medium \cite{3,4,5,6,7}.

There is a great interest in extending these studies to
semiconductors, not only for the understanding of the nature of
quantum coherence in semiconductors but also for the possible
implementation of optical devices such as XPM phase shifter
\cite{12}, switches \cite{13}, etc. It is well known, in the
conduction band of semiconductor quantum structure, that the
confined electron gas exhibits atomic-like properties. For example,
it has been shown that they can lead to gain without inversion
\cite{14,15,16}, coherently controlled photocurrent generation
\cite{17}, electron intersubband transmissions \cite{18}, and EIT
\cite{19,20}, slow light \cite{21}, interferences \cite{22}, optical
bistability \cite{0}, etc. Devices based on intersubband transitions
in SQW structures have many inherent advantages such as large
electric dipole moments due to the small effective electron mass,
high nonlinear optical coefficients, and a great flexibility in
device design by choosing the materials and structure dimensions.
Furthermore, the transition dipole energies can be controlled by
an external bias voltage. The implementation of XPM phase shift in
semiconductor-based devices is very attractive from a viewpoint of
applications, such as electro-optical modulators.

In this paper, we show the formation of ultra-slow bright and dark
solitons in semiconductor double quantum wells using intersubband
transitions by applications of a pulsed probe field and a continuous
wave (cw) strong control laser field. By choosing appropriate parameters, we also show the generation of a large XPM phase shift. As shown in Fig. 1, we consider a quantum well structure with three
energy levels that forms the well known cascade configuration
\cite{23}. $\omega_{21}$ and $\omega_{32}$ present the energy
differences of the $\left| 1 \right\rangle\leftrightarrow\left| 2
\right\rangle$ and $\left| 2 \right\rangle\leftrightarrow\left| 3
\right\rangle$, respectively. As a rule, such SQW samples are grown
by molecular beam epitaxy (MBE) method. The sample consists 30
periods, each with 4.8 nm
$\mbox{In}_{0.47}\mbox{Ga}_{0.53}\mbox{As}$, 0.2 nm
$\mbox{Al}_{0.48}\mbox{In}_{0.52}\mbox{As}$, and 4.8nm
$\mbox{In}_{0.47}\mbox{Ga}_{0.53}\mbox{As}$ coupled quantum wells,
separated by modulation-doped 36 nm
$\mbox{Al}_{0.48}\mbox{In}_{0.52}\mbox{As}$ barriers. The sample can
be designed to have desired transition energies, i.e., $E_{12}$ in
the range of 185 meV and $E_{23}$ in the range of 124 meV. Here, we
consider a transverse magnetic polarized probe incident at an angle
of 45 degrees with respect to the growth axis so that all transition
dipole moments include a factor $1/\sqrt{2}$ as intersubband
transitions are polarized along the growth axis. The sheet electron
density is about $4.7\times10^{11}\mbox{cm}^{-2}$. By using the
standard approach (this method has described quantitatively the
results of several literature \cite{13,15,16,18,20,24,25}), under
the rotating-wave and electro-dipole approximations the
semiclassical Hamiltonian describing the electron-field interaction
for the system under study in the Schr\"odinger picture, is given by

\begin{equation}
\label{eq1} H = \sum\limits_{j = 1}^3 {E_j \left| j \right\rangle
\left\langle j \right|} - \hbar (\Omega _c e^{ - i\theta _c }\left|
3 \right\rangle \left\langle 2 \right| + \Omega _p e^{ - i\theta _p
}\left| 2 \right\rangle \left\langle 1 \right| + h.c.),
\end{equation}

\begin{figure}
\includegraphics[width=12cm]{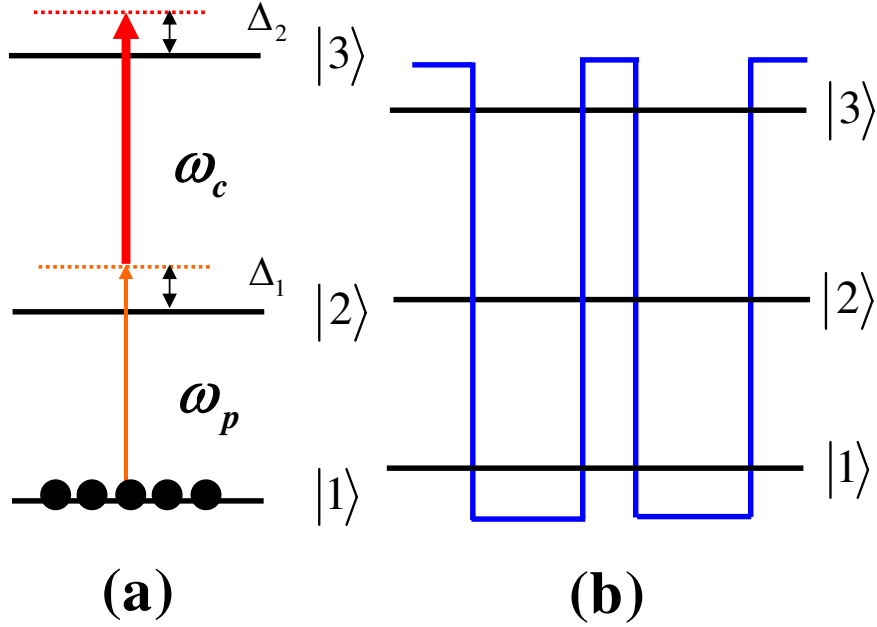} \caption{\textbf{(a)}Schematic energy level
arrangement for the quantum wells under consideration here. Subband
levels are labeled as $\left| 1 \right\rangle$, $\left| 2
\right\rangle$, and $\left| 3 \right\rangle$, respectively. The
subband transition $\left| 1 \right\rangle \leftrightarrow\left| 2
\right\rangle$ is driven by an weak probe field with central
frequency $\omega_p$ and the subband transition $\left| 2
\right\rangle \leftrightarrow\left| 3 \right\rangle$ is coupled by a
control field with central frequency $\omega_c$. \textbf{(b)}
Schematic three-level cascade electronic system synthesized
in a semiconductor quantum well.} \label{fig1}
\end{figure}

\noindent where the symbol h.c. means the Hermitian conjugate,
$\theta_n=k_n\cdot r -\omega_n t$ corresponds to the positive
frequency part of the respective optical field, $\Omega_n (n=p,c)$
are one-half Rabi frequencies for the relevant laser-driven
intersubband transitions, and $E_j=\hbar\omega_j (j=1-3)$ is the
energy of the subband $\left| j \right\rangle$. For simplicity, in
following analysis we will take $\omega_1=0$ for the ground-state
level $\left| 1 \right\rangle$ as the energy origin. Turning to the
interaction picture, with the assumption of $\hbar=1$, the free and
the interaction Hamiltonian can be respectively rewritten as follows

\begin{eqnarray}
\label{eq2} &&H_0 = \omega _p \left| 2 \right\rangle \left\langle 2
\right| + (\omega _p + \omega _c )\left| 3 \right\rangle
\left\langle 3 \right|,\\
\label{eq3} &&H_I = - \Delta _1 \left| 2 \right\rangle \left\langle
2 \right| - \Delta _2 \left| 3 \right\rangle \left\langle 3 \right|
- (\Omega _c e^{ik_c \cdot r}\left| 3 \right\rangle \left\langle 2
\right| + \Omega _p e^{ik_p \cdot r}\left| 2 \right\rangle
\left\langle 1 \right| + h.c.),
\end{eqnarray}

\noindent where the intersubband transition detunings of the two
optical fields are defined respectively by $\Delta_1=\omega_p-E_{2}$
and $\Delta_2=\omega_p+\omega_c-E_{3}$. Let us assume the electronic
wave function of the form

\begin{equation}
\label{eq4} \left| \psi \right\rangle = A_1 \left| 1 \right\rangle +
A_2 e^{ik_p \cdot r}\left| 2 \right\rangle + A_3 e^{i(k_p + k_c )
\cdot r}\left| 3 \right\rangle,
\end{equation}

\noindent together with $A_{j} (j=1,2,3)$ being the time-dependent
probability amplitudes of finding the electron in subbands $\left| j
\right\rangle$. By using the Schr\"odinger equation in the
interaction picture $i\partial\left| \psi \right\rangle/\partial
t=H_I\left| \psi \right\rangle$ for the three level model, the
equations of the motion for the probability amplitude of the
electronic wave functions and the wave equation for the
time-dependent probe field can be readily obtained as

\begin{eqnarray}
\label{eq5} &&\frac{\partial A_2}{\partial
t}=i(\Delta_1+i\gamma_2)A_2 + i\Omega_c^*A_3 + i\Omega_pA_1,\\
\label{eq6} &&\frac{\partial A_3}{\partial
t}=i(\Delta_2+i\gamma_3)A_3 + i\Omega_cA_2,\\
\label{eq7} &&\left| {A_1 } \right|^2 + \left| {A_2 } \right|^2 +
\left| {A_3 } \right|^2 = 1,\\
\label{eq8} &&\frac{\partial\Omega_p}{\partial z} +
\frac{1}{c}\frac{\partial\Omega_p}{\partial
t}=i\frac{2N\omega_p\left| {\mu_{21} } \right|^2}{c}A_2A_1^*,
\end{eqnarray}

\noindent with $N$ and $\mu_{12}$ being the concentration and the
dipole moment between states $\left| 1 \right\rangle$ and $\left| 2
\right\rangle$, respectively. In writing Eq. (\ref{eq8}), we have
assumed collinear propagation geometry and applied slowly varying
envelope approximation. $\gamma_2$ and $\gamma_3$ denote the total
decay rates of the subbands $\left| 2 \right\rangle$ and $\left| 3
\right\rangle$, which are added phenomenologically \cite{13,18} in
the above coupled amplitude equations. In semiconductor quantum
wells, the overall decay rate $\gamma_i$ of the subband $\left| i
\right\rangle$ comprises a population-decay contribution
$\gamma_{il}$ as well as a dephasing contribution $\gamma_{id}$,
i.e., $\gamma_i=\gamma_{il}+\gamma_{id}$. the former $\gamma_{il}$
is due to longitudinal optical (LO) photon emission events at low
temperature. The latter $\gamma_{id}$ may originate not only from
electron-electron scattering and electron-phonon scattering, but
also from inhomogeneous broadening due to the scattering on
interface roughness. The population decay rates can be calculated by
solving the effective mass Schr\"{o}dinger equation. For the
temperatures up to 10 K, the carrier density smaller than $10^{12}$
$\mbox{cm}^{-2}$, the dephasing decay rates $\gamma^{dph}_{ij}$ can
be estimated according to Ref.\cite{13}. For the SQW structure
considered here, the total decay rates turn out to be
$\gamma_2=\gamma_3=5 \mbox{meV}$. A more complete theoretical
treatment taking into account these processes for the dephasing
rates is though interesting but beyonds the scope of this paper.

In order to describe clearly the interplay between the dispersion
and nonlinear effects of the SQW system interacting with two optical
fields (probe and control fields), we now first focus on the
dispersion properties of the system. It requires perturbation of the
system respective to the first order of probe field $\Omega_p$ while
keeping full orders of control field $\Omega_c$. In the following,
we show effects from higher-order $\Omega_p$, and those required to balance the dispersion effect, resulting the
formation of ultraslow solitons. From Eqs. (\ref{eq5}-\ref{eq8}), it
is readily obtain that \cite{26,27}

\begin{equation}
\label{eq9} \frac{\partial\Omega_p}{\partial z} +
\frac{1}{c}\frac{\partial\Omega_p}{\partial t} =
iA_1^*[K(\hat{\omega})-\frac{\hat{\omega}}{c}](\Omega_pA_0),
\end{equation}

\noindent where $i\partial/\partial t$ is a differential operator
and with sufficiently intense control field we have

\begin{equation}
\label{eq10}
K(\hat{\omega})=\frac{\hat{\omega}}{c}+\frac{\epsilon_{12}(\hat{\omega}+\Delta_2+i\gamma_3)}{\left|
{\Omega_c }
\right|^2-(\hat{\omega}+\Delta_1+i\gamma_2)(\hat{\omega}+\Delta_2+i\gamma_3)}\simeq
K_0+\frac{\hat{\omega}}{v_g}+K_2\hat{\omega}^2 + O(\hat{\omega}^3),
\end{equation}

\noindent where $\epsilon_{12}=2N\omega_p\left| {\mu_{12} }
\right|^2/c$ and higher-order derivative terms have been neglected. The
physical interpretation of Eq. (\ref{eq10}) is rather clear. $K_0 =
\Phi + i\alpha $ describes the phase shift $\Phi $ per unit length
and the absorption coefficient $\alpha $ of the pulsed probe field,
$K_1$ gives the group velocity $V_g = \mbox{Re}[1 / K_1]$, and $K_2$
represents the group-velocity dispersion that contributes to the
probe pulse's shape change and additional loss of the pulsed probe
field intensity. With the dispersion coefficients obtained, then we
describe the nonlinear evolution of the probe field. We should
emphasize that it is indeed possible to obtain a set of
experimentally achievable parameters that lead to the
formation of ultraslow solitons, and solitons produced in this way
generally travel with a group velocity given by $V_g = \mbox{Re}[1 /
K'(0)]$. Considering the situation that almost all electrons will
remain in the subband level $\left| 1 \right\rangle$ due to the fact
that the laser-matter interaction is weak, we hence assume that $A_1(t=0)=1$ and the strong pump condition that the control
laser is strong enough to make $\kappa=\Omega_p/\Omega_c$ be a small
parameter (weak probe approximation). Then taking
$A_j=\Sigma_nA_j^{(n)}$ with $A_j^{(n)}=O(\kappa^n)$ and assuming
the adiabatic condition $\hat{\omega}/\Omega_c=O(\kappa)$, we have
the results

\begin{eqnarray}
\label{eq11}
&&A_1^*[K(\hat{\omega})-\frac{\hat{\omega}}{c}](\Omega_pA_1)=\left|
{A_1 }
\right|^2[K(\hat{\omega})-\frac{\hat{\omega}}{c}]\Omega_p+O(\kappa^4),\\
\label{eq12}
&&K(\hat{\omega})\Omega_p=[K_0+\frac{\hat{\omega}}{v_g}+K_2\hat{\omega}^2]\Omega_p+O(\kappa^4),\\
\label{eq13} &&\left| {A_1 } \right|^2 =1 - \left| {A_2 } \right|^2
+ \left| {A_3 } \right|^2,
\end{eqnarray}

\noindent with $A_j$, (j=2,3,4), given by

\begin{equation}
\label{eq14}
A_j=\frac{[(\Delta_2+i\gamma_3)\delta_{j2}-\Omega_c^*\delta_{j1}]\Omega_p}{\left|
{\Omega_c } \right|^2-(\Delta_1+i\gamma_2)(\Delta_2+i\gamma_3)} +
O(\kappa^2).
\end{equation}

\noindent Equation (\ref{eq14}) is readily obtained by solving Eqs.
(\ref{eq5}-\ref{eq6}) under the steady state condition, i.e., $\partial
A_{2,3}/\partial t =0$ and $A_1^{(1)}=1$. Here we have used the
relations $\partial A_{2,3}/\partial t=O(\kappa
A_{2,3})=O(\kappa^2)$ and $A_1=1+O(\kappa^2)$. Substituting
$\Omega_p(z,t)=\Omega_p(z,t)\mbox{exp}(iK_0z)$ into Eq. (\ref{eq9})
and using above results and discussion, it is then
straightforward to obtain the following nonlinear evolution
equation, which is accurate up to the order $O(\kappa^3)$, for the
slowly-varying envelope $\Omega_p (z,t)$,

\begin{equation}
\label{eq15}
i\frac{\partial\Omega_p}{\partial\xi}-K_2\frac{\partial^2\Omega_p}{\partial\eta^2}=
We^{-\alpha\xi}\left| {\Omega_p } \right|^2\Omega_p,
\end{equation}

\noindent here we have assumed $\xi=z$, $\eta=t-z/v_g$. The velocity
$v_g$ and the dispersion coefficient $K_2$ are determined by Eq.
(\ref{eq10}), the absorption coefficient $\alpha=\mbox{Im}(K_0)$ and
the nonlinear coefficient $W$ are explicitly given by

\begin{eqnarray}
\label{eq16}
&&\alpha=\mbox{Im}[\frac{\epsilon_{12}(\Delta_2+i\gamma_3)}{\left|
{\Omega_c } \right|^2-(\Delta_1+i\gamma_2)(\Delta_2+i\gamma_3)}],\\
\label{eq17} &&W=\frac{\epsilon_{12}(\Delta_2+i\gamma_3)(\left|
{\Omega_c } \right|^2+\Delta_2^2)+\gamma_3^2}{[\left| {\Omega_c }
\right|^2-(\Delta_1+i\gamma_2)(\Delta_2+i\gamma_3)]\left| {[\left|
{\Omega_c }
\right|^2-(\Delta_1+i\gamma_2)(\Delta_2+i\gamma_3)]}\right|^2}.
\end{eqnarray}

Now we briefly discuss the cross-phase modulation (XPM). Let us
consider the following parameter condition: $\Delta_1\simeq0$, with
other parameters unchanged and writing
$K_{0}L=\Phi_{\mbox{xpm}}+i\alpha L$ ($L$ is the length of the SQW
system), it is straightforward to show that

\begin{equation}
\label{eq00} \Phi_{\mbox{xpm}}\simeq\frac{\left| {\Omega_c }
\right|^2\Delta_2\epsilon_{12}}{\gamma_2^2\Delta_2^2+(\left|
{\Omega_c } \right|^2+\gamma_2\gamma_3)^2}, \alpha\simeq\frac{\left|
{\Omega_c }
\right|^2\gamma_3\epsilon_{12}}{\gamma_2^2\Delta_2^2+(\left|
{\Omega_c } \right|^2+\gamma_2\gamma_3)^2}.
\end{equation}

\noindent These results in our structure are similar to those of the
giant cross-phase modulation in cold atom media \cite{4}, but, we
only need one control laser field and do not need to introduce a second
control laser field. The ratio of $\Phi_{\mbox{xpm}}/\alpha L$,
characterizing the ability achieving the cross-phase modulation
phase shift without appreciated absorptions, has the form
$\Delta_2/\gamma_3$ and is independent of the coupling field
intensity. Furthermore, since the intersubband energy level can be
easily tuned by an external bias voltage, thus we may provide
another possibility to realize electrically controlled phase
modulator at low light levels.

If a reasonable and realistic set of parameters can be found so that
$\mbox{exp}(-\alpha L) \simeq 1$, i.e., the losses of the probe
pulse are small enough to be neglected, thus the balance between the
nonlinear self-phase modulation and the group velocity dispersion
(described by the coefficient $K_2$) may keep a pulse with
shape-invariant propagation, which yields $K_2 = K_{2r} + iK_{2i}
\simeq K_{2r}$, and $W = W_r + iW_i \simeq W_r$. Then Eq.
(\ref{eq15}) can be reduced to the standard nonlinear
Schr\"{o}dinger equation governing the pulsed probe field evolution
\cite{3,4}

\begin{equation}
\label{eq18}
i\frac{\partial\Omega_p}{\partial\xi}-K_{2r}\frac{\partial^2\Omega_p}{\partial\eta^2}=W_r\left|
{\Omega_p } \right|^2\Omega_p,
\end{equation}

\noindent which admits of solutions describing bright
($K_{2r}W_r<0$) and dark ($K_{2r}W_r>0$) solitons, including
$N$-soliton ($N=1,2,3,\ldots$) solution for dark and bright solitons. And
whether the solutions to Eq. (\ref{eq18}) are the bright or
dark solitons depends on the sign of product $K_{2r}\cdot W_r$.
The single soliton is called as the fundamental soliton, and
$N$-soliton ($N=2,3,\ldots$) is named as the higher-order soliton.

The fundamental dark soliton of Eq. (\ref{eq18}) with $K_{2r}W_r<0$
is

\begin{equation}
\label{eq19} \Omega _p = \Omega _{p0} \mbox{tanh}(\eta / \tau )\exp
[-i\xi W_r\left| {\Omega _{p0} } \right|^2],
\end{equation}

\noindent where amplitude $\Omega _{p0} $ and width $\tau $ are
arbitrary constants subjected only to the constraint $\left| {\Omega
_{p0} \tau } \right|^2 = - 2K_{2r}/W_r$.

The fundamental bright soliton, and the bright 2-soliton (bright
second-order soliton) of Eq. (\ref{eq18}) with $K_{2r}W_r>0$ are
given respectively by

\begin{eqnarray}
\label{eq20} &&\Omega _p = \Omega _{p0} \mbox{sech}(\eta / \tau
)\exp
[-i\xi W_r\left| {\Omega _{p0} } \right|^2/2],\\
\label{eq21} &&\Omega _p = \Omega _{p0}
\frac{4[\mbox{cosh}(3\eta/\tau)+3\mbox{exp}(-8iK_{2r}\xi/\tau^2)\mbox{cosh}
(\eta/\tau)]\mbox{exp}(-iK_{2r}\xi/\tau^2)}{\mbox{cosh}(4\eta/\tau)+4\mbox{cosh}(2\eta/\tau)+3\mbox{cos}(8K_{2r}\xi/\tau^2)},
\end{eqnarray}

\noindent where the amplitude $\Omega _{p0} $ and width $\tau $ are
arbitrary constants subjected only to the constraint $\left| {\Omega
_{p0} \tau } \right|^2 = 2K_{2r}/W_r$. It is worth to note that the
bright 2-soliton solution in Eq. (\ref{eq21}) satisfies $\Omega_p
(\xi=0,\eta)=2\Omega_{p0}\mbox{sech}(\eta/\tau)$.

\begin{figure}\label{fig2}
\includegraphics[width=12cm]{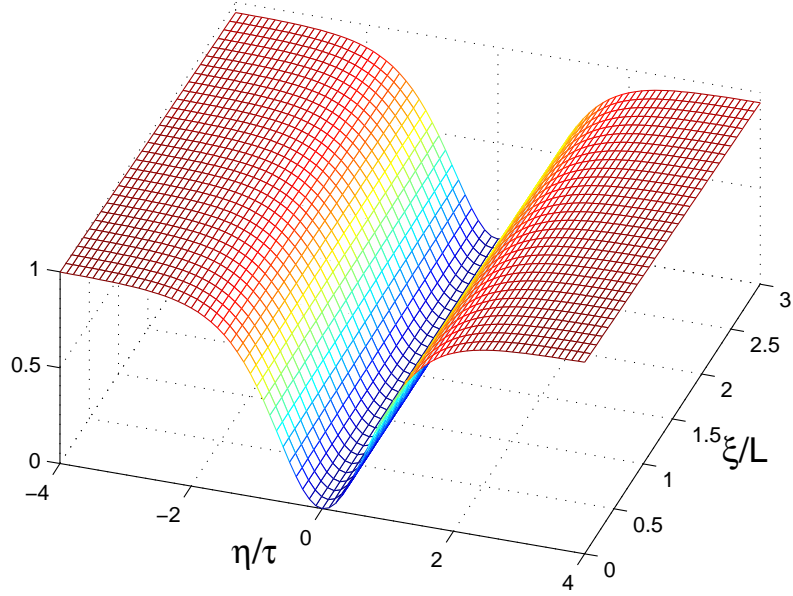}
\caption{Surface plot of the amplitude for the generated fundamental
dark soliton
$\left|\Omega_{p}/\Omega_{p0}\right|^2\mbox{exp}(-2\alpha\xi)$ with
$\left|\Omega_{p}\right|^2$ being the numerical solution to Eq.
(\ref{eq15}) versus dimensionless time $\eta/\tau$ and distance
$\xi/L$ under the initial condition $\Omega_{p}(\xi = 0,\eta) =
\Omega_{p0}\mbox{tanh}^2(\eta/\tau)$, where $L=1.0$ cm, $\tau=1.0\times10^{-6}$s, and other simulation parameters are explained in the main text.} \label{fig2}
\end{figure}

Our scheme is different from EIT in a SQW structure, in which the
latter can not form solitons. Because that slow group velocity
propagation requires weak driving conditions, this leads to very
narrow transparency windows. Thus EIT operation with weak driving
conditions requires single and two-photon resonance excitations,
i.e., $\Delta_1=\Delta_2=0$ in Eq. (\ref{eq17}). Deviations from
these conditions will result in significant probe field attenuation
and distortion. Besides, one can find that the nonlinear coefficient $W$
is almost purely imaginary under these EIT conditions. This is
contradictory to the requirement of $W\simeq W_r$ in order to
preserve the complete integrability of the standard nonlinear
Schr\"odinger Eq. (\ref{eq18}). However, here we have found that by
appropriately choosing the intensities and detunings of laser
fields, we can achieve $\mbox{exp}(-\alpha L)\simeq1$ for $L$ within
a few centimeters, $K_2\simeq K_2r$, $W\simeq W_r$, and ultraslow
group velocities for both bright and dark solitons studied in this
work with the typical population decay and dephasing decay rates of
the transitions in SQW structures. Considering a system where the
total decay rates are $\gamma_2=\gamma_3=5 \mbox{meV}$, the parameters used are typical values for
transitions  $\left| 1 \right\rangle\leftrightarrow\left| 2
\right\rangle$ and $\left| 2 \right\rangle\leftrightarrow\left| 3
\right\rangle$ in SQW structures.

As an example, we now present numerical examples to demonstrate the
existence of ultraslow dark solitons in the system studied through
simulating the Eq. (\ref{eq15}) with the initial condition
$\Omega_p(\xi=0, \eta)=\Omega_{p0}\mbox{tanh}(\eta/\tau)$. Take
$\epsilon_{12}=80 \mbox{cm}^{-1} \mbox{meV}$, $\Omega_c=8
\mbox{meV}$, $\Delta_1=-10\mbox{meV}$, $\Delta_2\simeq0$, and
$\gamma_2=\gamma_3=5 \mbox{meV}$, we have
$V_g/c\simeq2.7\times10^{-4}$, and
$\alpha\simeq0.00019\mbox{cm}^{-1}$. With these parameters, the
standard nonlinear Schrodinger equation (\ref{eq18}) with
$K_{2r}\cdot W_r<0$ is well characterized, and thus we have
demonstrated that the existence of dark solitons that travel with
ultraslow group velocities in SQW structures. As shown in Fig. 2,
the numerical simulation of Eq. (\ref{eq15}) for the fundamental
dark soliton shows an excellent agreement with Eq. (\ref{eq19}).

It is worth to note that all the parameter sets also lead to
negligible loss of the probe field for both the bright and dark
solitons (including 2-soliton) described here. Besides, we have used the
one-dimensional model in calculation where the momentum-dependency
of subband energies has been ignored. However, there is no large
discrepancy between the reduced one-dimensional calculation
\cite{13} and the full two-dimensional calculation \cite{19,28}.

In conclusion, using the coupled Schr\"odinger-Maxwell equations for
a three-level system of electronic subbands, we have presented and
analyzed a novel scheme to achieve ultraslow bright and dark optical
solitons, and a large XPM phase shift can also be obtained with
appropriate parameters. Such investigations of ultraslow optical
solitons in the present work may lead to important applications including high-fidelity optical delay lines and optical buffers in SQW structures. Besides, a large XPM phase shift achieved in our proposed SQW
structure may open up an avenue to explore possibilities for
nonlinear optics and quantum information processing in a solid-state system
and may result in substantial impacts on technology of electrically
controlled phase modulator.

\acknowledgments The research is supported in part by National
Natural Science Foundation of China under Grant Nos. 10704017,
10634060, 90503010 and 10575040, by National Fundamental Research
Program of China 2005CB724508.

\end{document}